\documentstyle[12pt,twocolumn]{article}

\newcommand{\BEQ}{\begin{equation}}
\newcommand{\BEA}{\begin{eqnarray}}
\newcommand{\EEQ}{\end{equation}}
\newcommand{\EEA}{\end{eqnarray}}

\newcommand{\lm}{\lambda}
\newcommand{\lc}{\lambda_{c}}
\newcommand{\D}{\Delta}
\newcommand{\rar}{\rightarrow}
\newcommand{\tgi}{t \rightarrow \infty}
\newcommand{\rb}{\overline {\rho}}

\newcommand{\lra}{\leftrightarrow}

\begin{document}
\begin{titlepage}
\begin{center}
{\bf Generalized Scaling for Models with Multiple Absorbing States} \\
\vskip .5in
{\bf J. F. F. Mendes} \footnote{Permanent address: Departamento
de F\'\i sica da Universidade do Porto, Pra\c ca Gomes Teixeira,
4000 Porto - Portugal.\\
Supported partially by JNICT, Junta Nacional de
 Investiga\c c\~ao Cient\'\i fica e Tecnol\'ogica, project CEN-386/90
and by a PhD grant from JNICT.}
\footnote{Address after 23 October 1993: Department of Physics,
Theoretical Physics,
University of Oxford, $1$ Keble Road, Oxford OX$1$ $3$NP, UK },
{\bf Ronald Dickman}
\vskip 0.15in
Department of Physics and Astronomy, Herbert H. Lehman College, \\
 City University of New York, Bronx, New York $10468$
\vskip 0.2in
{\bf Malte Henkel}
\vskip 0.15in
Department of Physics, Theoretical Physics, University of Oxford, \\
1 Keble Road, Oxford OX1 3NP, UK
\vskip 0.15in
{\bf M. Ceu Marques}
\vskip 0.15in
Departamento de F\'\i sica da Universidade do Porto, Pra\c ca Gomes Teixeira,
\\
4000 Porto, Portugal
\vskip .5in
{PACS numbers: 64.60.-i, 05.50.+q, 05.70.Ce, 75.10.Hk } \\

\vskip 1.0cm
\begin{abstract}
At a continuous transition into a nonunique absorbing state,
particle systems may exhibit nonuniversal critical behavior,
in apparent violation of hyperscaling.
We propose a generalized scaling theory for dynamic critical
behavior at a transition into an absorbing state, which is capable of
describing exponents which vary according to the initial configuration.
The resulting hyperscaling relation is supported by simulations of
two lattice models.

\end{abstract}
\end{center}
\end{titlepage}

Nonequilibrium phase transitions continue to elicit great
interest in the physical
and biological sciences.
In hopes of better understanding nonequilibrium critical phenomena,
models exhibiting phase transitions into an
absorbing state are under intensive study in statistical physics.
The dichotomy between an absorbing (dead, inactive) state and an active
one arises naturally in such diverse areas as catalysis \cite{ziff86},
epidemiology \cite{harris74,liggett85,durrett},  and the transition to
turbulence \cite{pomeau}.

The essential features of the phase transition
are typified by  the contact process (CP) \cite{harris74}.
In the CP, each site of the lattice
{\bf Z}$^{d}$ is either occupied or vacant.
Occupied sites become vacant at unit rate, whilst a vacant site $i$
becomes occupied at rate $\lm q_{i}$, with $q_{i}$  the fraction of
occupied nearest neighbors of $i$.  Evidently, the vacuum is absorbing.  The
growth rate $\lm $ determines the ultimate survival of the population:
for $\lm < \lc$ the vacuum is the unique steady state, but for $\lm > \lc
(\simeq 3.298$ in one dimension), there is also an active stationary state
characterized by a nonzero particle density $\rb \propto (\lm - \lc)^{\beta}$.
The transition at $\lc $ is a nonequilibrium critical point, belonging
to the universality class of directed percolation (DP)
\cite{durrett,blease,kinzel} and Reggeon field theory \cite{cardy80}.
Indeed, studies of a host of models provide ample support for
the conjecture that continuous transitions into a unique absorbing state
generically fall in the DP class \cite{gra82,janssen}.

The situation regarding models with multiple absorbing states is more
complex.
On one hand, studies of some two-dimensional  catalysis
models yield critical exponents different from
those of DP \cite{didi,ditri,yald93}.
On the other,  the one-dimensional pair contact process (PCP)
and dimer reaction (DR)
clearly fall in the DP class, as far as {\it static} critical behavior
is concerned \cite{iwanpcp,infabs}.  (In all of these models, the number
of absorbing configurations grows exponentially with lattice size.)
The {\it dynamic} critical properties of the PCP and DR were found,
surprisingly, to be nonuniversal, the exponents depending upon the
nature of the initial configuration \cite{infabs}.  While this variation is
quite regular, it appears to violate a basic hyperscaling relation
amongst the exponents $\delta $, $ \eta, $ and $z$ (defined below),
suggesting a breakdown of the well-established scaling theory.

This apparent breakdown has prompted us to reexamine
the scaling hypothesis for models with multiple absorbing states.
We arrive at a scaling theory in which additional
exponents are expected to depend upon the starting configuration,
and in which the exponents satisfy a generalized hyperscaling
relation.  The latter is verified in
simulations of the DR and of a new
model called the {\it threshold transfer process}.
In what follows we define the models, present the
scaling theory, and report the numerical evidence supporting it.

Simulations of  the threshold transfer process (TTP)
permit us to study nonuniversal critical spreading in the
context of a three-state model, providing a check on
the robustness of earlier findings \cite{infabs}.
In the TTP each site may be
vacant, or singly or doubly occupied, corresponding to
$\sigma_i = 0, 1 $ or 2.
In each cycle of the evolution, a site
$i$ is  chosen at random. If $\sigma_i=0$,
then $\sigma_i \rightarrow 1 $ with probability $r$;
if $\sigma_i=1$, then $\sigma_i \rightarrow 0 $ with probability $1-r$.
(``0" and ``1" sites are left unchanged with probabilities $1-r$ and $r$,
respectively.)  In the absence of doubly-occupied sites, we have a trivial
dynamics in which a fraction $r$ of the sites have $\sigma_i =1$,
in the steady state.
However, if  $\sigma_i=2$, the particles may move
to neighboring sites.  If $\sigma_{i-1} < 2$, one particle moves to
that site; independently, a particle moves from $i$ to $i+1$ if
$\sigma_{i+1} < 2$.  $\sigma_i $ is diminished accordingly in this
deterministic, particle-conserving transfer.  Survival of the
doubly-occupied sites hinges on the process
$(1,2,1) \rightarrow (2,0,2)$ (their number decreases or
remains the same in all other events), and so
depends upon the parameter $r$, which controls the particle
density.  The set of configurations devoid of doubly-occupied
sites comprises an absorbing subspace of the dynamics, which
can be avoided only if $r$ is
sufficiently large.   Thus
we identify
the density of doubly-occupied sites, $\rho_2$, as the order
parameter of the threshold transfer process.

We note in passing  that the TTP
bears some resemblance to a sandpile model devised by
Manna \cite{manna91}, and to a
forest fire model (FFM) proposed
by Bak, Tang and Weisenfeld \cite{bak87,gras91}.
Under the correspondence: $2 \lra $ burning tree, $1 \lra$ live tree, and
$0 \lra$ ashes, the process ${1,2,1} \rightarrow {2,0,2} $ describes
a burning tree setting its neighbors on fire, and $r$ represents the rate
at which new trees emerge from the ashes. But the TTP
permits doubly occupied sites to arise only {\it via } transfer; there is
no  ``lightning" process, as in the FFM.  A further contrast is that
a rule such as ${0,2,0} \rar {1,0,1}$ has no analog
in the FFM.
Despite certain common aspects, our model is therefore very different from
the FFM.

The dimer reaction (DR), introduced in Ref. \cite{infabs}, is a lattice model
in which sites may be either vacant
or (singly) occupied;
particles may not occupy adjacent sites.
In each step of the process
a site $i$ is selected at random.
If $i$ is occupied, or is blocked by a neighboring
particle then nothing happens.  But if
$i$ is {\it open} (i.e., $i$ and its neighbors
are vacant), a new particle appears, which may
remain at $i$, or react with another
particle, depending on the occupancy of the nearby sites:

$(i)$ If at least one of the second neighbors, $i-2$, and $i+2$, are
occupied, then with probability $1-p$
there is a reaction between the new particle and its neighbor
(chosen at random if there is a choice), which removes them both;
with probability $p$ there is no reaction and the new particle remains.

$(ii)$ If both second neighbors
are vacant, but at least one of the third neighbors is occupied, then
a reaction with a third-neighbor particle may occur, with
probabilities as in case $(i)$.

$(iii)$ If none of the second or third neighbors are occupied, the
new particle remains at site $i$.

\noindent In the DR, any configuration devoid of open sites is
absorbing.  The order parameter is the stationary open site
fraction, $\rb_o$.

Both the TTP and the DR exhibit continuous phase transitions to an absorbing
state marked by a vanishing order parameter
at a critical value of $r$ or of $p$.   In this work
we are concerned with {\it critical spreading}, that is, the
evolution of a critical system from a nearly-absorbing
initial configuration.
The exponents describing this spreading are nonuniversal, i.e.,
they depend upon the particle density in the initial state \cite{infabs}.
Before reporting our numerical results, we present a scaling theory for
such processes.

Following Grassberger and de la Torre \cite{torre}, we consider
an ensemble of trials, all starting from the same initial
configuration: a single seed  in an otherwise absorbing configuration.
(For the contact process, this means one particle in an
otherwise empty lattice; for the DR, one open site; for the
TTP, one doubly-occupied site.)  Let $\rho (x,t) $
denote the local order-parameter density,
$\D $  the distance from the critical
point ($\D = \lm - \lc $ in the contact process).
In the critical region the system is characterized by
a correlation length $ \xi_{\perp} \propto \D^{-\nu_{\perp}} $,
and relaxation time $ \tau \propto \D^{-\nu_{||}} $.
At the critical point the asymptotic evolution is described by
power-laws; for $\D \neq 0$ the power laws are modified by
scaling functions which depend upon the dimensionless ratios
$x/\xi_{\perp} $ and $t/\tau$.
Thus the survival probability - i.e., that a trial has
evaded the absorbing state, $\rho (x) \equiv 0$ -
is expected to follow
\BEQ
\label{survsc}
P(t) \simeq t^{-\delta} \phi(\Delta t^{1/\nu_{||}}),
\end{equation}
so that $P \propto t^{-\delta} $ at the critical point.
The order parameter density (averaged over all trials) is
\BEQ
\label{densc}
\rho(x,t) \simeq t^{\eta - dz/2} F(x^{2}/t^{z}, \Delta t^{1/\nu_{||}}),
\EEQ
where the  $x$-dependence reflects symmetry and power-law
critical spreading from the seed at $x=0$.  For $\D =0$, one finds upon
integrating Eq (\ref{densc})  over space,
that the mean population
$n(t) \propto t^{\eta} $, whilst
the second moment  implies that  the
mean-square spread of the population
$ R^2 (t) = \; \langle x^2 \rangle _t \;\; \propto
t^z $.  The exponents $\delta $, $ \eta $ and $z$ characterize
critical spreading;
several relations connect them with other exponents.

Consider first the CP,  for which the ultimate survival probability,
$P_{\infty} \equiv \lim_{\tgi} P(t) = \rb$,
the stationary particle density \cite{torre}.
Existence of the limit requires
$\phi (x) \propto x^{\delta / \nu_{||} } $ for large
$x$, and $\rb \propto \D^{\beta}$ then implies the scaling relation
\BEQ
\delta = \beta/\nu_{||}.
\EEQ
For $\Delta < 0$ and (large) fixed $t$ we expect $\rho(x,t) \simeq
e^{-x/\xi}$, which implies (since
$\xi \propto \Delta^{-\nu_{\perp}}$), that for $v < 0$,
$F(u,v) \propto \exp(-const. \sqrt {u} |v|^{\nu_{\perp}})$.
In order that $\xi$ be time-independent, we must have
\begin{equation}
z = 2 \nu_{\perp}/\nu_{||}.
\end{equation}
\noindent Finally, note that for $\D > 0$,  the local density at any
fixed $x$, in a surviving trial,
must approach $\rb$ as $\tgi$. Since
$\rho(x,t)$  represents an average over {\em all} trials, we have
\begin{equation}
\rho(x,t) \rightarrow P_{\infty}\Delta^{\beta} \propto \Delta^{2\beta},
\end{equation}
\noindent as $t \rightarrow \infty$,
which implies that $F(0,v) \propto v^{2\beta}$
for large $v$.  Existence of a stationary state then
requires that the exponents satisfy the hyperscaling relation
\begin{equation}
\label{hyper}
4 \delta + 2 \eta = dz.
\end{equation}

We turn now to models such as the pair contact process (PCP) \cite{iwanpcp},
the dimer reaction \cite{infabs}, and the threshold transfer process,
which possess a multitude of absorbing configurations.
Absorbing configurations in the PCP and DR can have various particle
densities; the analogous variable in the TTP is the density $\rho_1 $
of singly occupied sites.  We refer to this aspect of the
(near-absorbing) initial configuration in a critical spreading process
as the ``initial density,"
$\phi_i$.  One value of $\phi_i$ is special in these models: the ``natural"
particle density $\phi_c$ of the quasistationary critical process.
(For the DR, $\phi_c \simeq 0.418$; for the TTP, $\phi_c \simeq 0.69$.)
Simulations
indicate that in each of these one-dimensional models, the {\it static}
critical behavior belongs to the directed percolation class, but that the
exponents $\delta $, $\eta $ and $z$ are {\it nonuniversal}, varying
continuously with initial density.   (The critical point $\lc $, by contrast,
does not change as $\phi_i$ is varied.)
Only when $\phi_i = \phi_c$ do the exponents
assume DP values, and only then do they satisfy Eq (\ref {hyper}).
Rather than interpreting this as a violation of hyperscaling, we shall
argue that in these models the scaling hypothesis must be modified,
leading to a generalized hyperscaling relation.

We assume that as in models with a unique absorbing state,
the order parameter density has the scaling form
\begin{equation}
\rho(x,t) \simeq t^{\eta' - dz'/2} G(x^{2}/t^{z'}, \Delta t^{1/\nu_{||}'}),
\end{equation}
where the primed exponents are functions of $\phi_i$. Similarly,
we suppose the survival probability follows
\begin{equation}
\label{gensp}
P(t) \simeq t^{-\delta'} \Phi(\Delta t^{1/\nu_{||}'}).
\end{equation}
Since the stationary distribution is unique, we have as before
that
\begin{equation}
\rho(x,t) \rightarrow P_{\infty}(\phi_i) \Delta^{\beta},
\end{equation}
as $\tgi$, with $\beta $ the usual directed percolation exponent.  However
there is no reason to suppose that
$P_{\infty} (\phi_i) \propto \overline {\rho}$,
when $\phi_i \neq \phi_c$.  In fact if this were so,
we would have $\delta' \nu_{||}' = \beta$, implying that the
primed exponents satisfy Eq(\ref{hyper}).  Since they do not, we
conclude that the exponent  governing the ultimate
survival probability must also depend upon $\phi_i$, i.e.,
$ P_{\infty} \propto \Delta^{\beta'} $, with
$ \beta' = \delta' \nu_{||} '$.
By the same arguments applied to the CP we find
\BEQ
\label{gzsc}
z' = 2 \nu_{\perp}'/\nu_{||}',
\EEQ
where we have introduced exponents $\nu_{||}'$ and $\nu_{\perp}'$
which govern the mean lifetime and spatial extent of a cluster grown
from a single seed.  The asymptotic behavior of the order parameter density
is now  $\rho(x,t) \rightarrow \Delta^{\beta + \beta'}$,
and $G(0,y) \propto y^{\beta + \beta'} $ for large $y$, which implies
the generalized hyperscaling relation
\begin{equation}
\label{ghyper}
2(1 + \frac{\beta}{\beta'}) \delta' + 2 \eta' = dz'.
\end{equation}

We have verified Eq (\ref{ghyper}) in simulations of the dimer reaction
and the threshold transfer process.
Using time-dependent simulations (for $t \leq 2000 $, and samples of
$10^5 $ to $8 \times 10^5$ trials),
 we determined the critical point
of the TTP as $r_{c}=0.6894(3)$.
Analysis of steady-state data for $\rb_2$, as shown
in figure  1, then yields $\beta=0.279(5)$,  in good agreement with the
value for DP in $1+1$ dimensions,  $\beta=0.2769(2)$
\cite{tdppt,gencp}.
The exponents $\delta '$, $\eta '$ and $z'$
may be determined from simulations at $r_c$, using an
initial configuration
very close to the absorbing state. We studied various
initial densities, including $\phi_i = 0.69 $,
the natural value.
The ultimate survival probability exponent $\beta '$ was determined from
similar studies, using $r$ values slightly above critical.
The simulations begin with one doubly-occupied site
at the origin; the remaining sites are taken as occupied or vacant,
independently, with probabilities $\phi_i $ and $1- \phi_i$, respectively.
The dynamics is restricted to an active region
defined as follows.  Let $\Lambda_i $ be the set of all sites which are
doubly-occupied or have a doubly-occupied neighbor, after
the $i^{\rm th}$ step
of the trial.  ($\Lambda_0  $ comprises the origin and its neighbors.)
The site to be updated at step $i+1$ is selected at random from
$\cup_{j=0}^{i} \; \Lambda_j $. Thus the evolution proceeds on an expanding
set within the ``light-cone" emanating from the origin.
As in the DR and the PCP, distant sites are not
updated until the active region reaches their neighborhood.
Figure  2 shows a local-slope analysis
for $\delta $, i.e., a plot of
$\delta(t) \equiv  \ln [P(mt)/P(t)]/\ln m $ {\it vs.}
$t^{-1}$, for various initial densities.  (In this study we used $m=5$.)
In  figure  3 we show typical results for the
ultimate survival probability, leading to an estimate for $\beta '$.

The simulation procedure for the DR is described in Ref. \cite{infabs}.
On the basis of  more extensive studies of the half-life $\tau $,
on lattices of up to 1000 sites, we now find
$p_c = 0.26401(2)$, consistent with the earlier result of $0.26400(5)$.
We used $p$ values slightly below critical ($p_c - p \leq 0.05$)  in
determinations of $\beta '$ at the four initial densities studied in
Ref.\cite{infabs}.     According to Eq \ref{gensp}, in a plot of
$\tilde{P} \equiv \D ^{- \beta '} P $
{\it vs.} $\tilde{t} \equiv \D t^{1/\nu_{||}'} $,
data for various $\D$ (for a particular $\phi_i$),
should fall on a single curve.
Figure 4 shows a reasonably good collapse of the data for each of
four initial densities.

Our results for the exponents in the TTP and DR are given in
Table I, together with a test of the new hyperscaling
relation, Eq (\ref{ghyper}).
Evidently it is confirmed to within the precision of the data. (By contrast,
the DP hyperscaling relation, Eq (\ref{hyper}), is clearly violated.) Thus
the spreading and survival exponents for transitions into a
nonunique absorbing state may be described using the conventional
sort of scaling theory, properly generalized to allow for a dependence upon
the initial density.

The dependence of $\delta '$, $\eta '$ and $\beta '$ upon initial density is
quite pronounced, that of $z$ much weaker.  We have made no determination
of $\nu_{\perp}'$, and our results for $\nu_{||}'$, which come solely from the
relation $\nu_{||}' = \beta '/\delta '$,
show no significant variation with $\phi_i$.
(We find $\nu_{||} = 1.80(6) $ and $ 1.76(6)$
for the TTP and the DR, respectively;
the DP value is $1.74(1) $.)  In light of Eq (\ref{gzsc}), it appears that
$\nu_{\perp}$ is not strongly dependent upon
initial density either.

Further examination of the data indicates that $\delta' + \eta'$ is also
very nearly constant.  This is clear
from the plot $\eta '$ {\it vs. } $\delta '$ for all
three models (TTP, DR, and PCP) shown in
figure 5.  (The slope of the linear best-fit
is -0.995.)  It is also worth noting that the exponents of the
(two-dimensional)
dimer-trimer model \cite{ditri}  differ
from those of DP, but that $\delta + \eta $
is again the same as in DP.  (Simulations of the dimer-trimer model yield
$\delta = 0.40(1)  $, $\eta = 0.28(1)  $,
compared with 0.460(6) and 0.214(8), respectively,
for two-dimensional DP \cite{gras89}.)
Now $\delta' + \eta' $ is the exponent governing the population growth
in {\it surviving} critical trials.  Its independence of $\phi_i$ suggests that
the asymptotic properties of a surviving
trial are not affected by the initial
density.  This conclusion is strongly supported
by the absence of any detectable shift in the critical point
as $\phi_i$ is varied.  As further support, one may note that as $\tgi$, only a
negligible fraction of a surviving cluster is
actually in contact with the external
density $\phi_i$.  Deep inside the cluster, the particle density must approach
the natural value $\phi_c$.
This point of view also implies that $z'$, which describes
surviving trials exclusively, should be constant.
In fact, if  $z'$ and $\delta' + \eta'$ are
constant, then Eqs (\ref{gzsc}) and (\ref{ghyper}) require that $\nu_{||} '$
and
$\nu_{\perp}'$ are as well.  We are led, by this line of argument, to a more
economical description of critical spreading,
in which all of the exponent variation
follows from a single cause: the dependence of the survival probability
upon initial density.  The ensuing predictions regarding exponents are
consistent with our numerical results, except for a small variation in $z'$
with
$\phi_i$.  One may argue, however, that the results for $z'$ are affected by
$\phi_i$-dependent corrections to scaling, and that a more precise
numerical test is needed.
In summary, we believe that the most natural
and parsimonious interpretation is that the initial density influences the
survival probability, but not the scaling properties of surviving events.

The pair contact process, dimer reaction, and threshold transfer
process all involve a second variable, $\phi $, dynamically
coupled to the order parameter.
A quantitative theory
of the dependence survival probability, and the associated
exponents $\delta ' $ and $\beta '$, upon the
initial density $\phi_i $ has yet to be devised.
But we can offer some intuitive basis for understanding
nonuniversality by suggesting that
in these processes, the
initial density plays a role analogous to that of a
{\it marginal parameter}. Such
parameters, invariant under renormalization group
transformations, often give rise to exponents which vary continuously along a
line of fixed points.   In the present case, $\phi_i$ represents a property of
the medium into which the process grows, and which is never forgotten, since
to survive, a critical process must repeatedly  invade new territory.
A RG transformation generally
involves coarse graining (which conserves density in the large),
and rescaling.  Such a transformation may be expected to
leave $\phi_i$, the density outside the active region, invariant,
while driving the correlation length of this region to zero.
Indeed, the spreading exponents are insensitive to (short
range) correlations in the initial state \cite{infabs}.
A more detailed understanding of nonuniversality in these
models may emerge when a suitable renormalization
group scheme is devised.
\vspace{6em}

\noindent {\bf Acknowledgements}

We thank Michel Droz, Iwan Jensen, and Eduardo Lage
for helpful discussions.  R.D. thanks the
Department of Theoretical Physics of the University of Geneva,
where a portion of this work was done,
for the warm hospitality he enjoyed during his visit. M.H. is grateful
to the Department of Theoretical Physics at the University of Geneva
and to the SERC for financial support.  The simulations were
performed on the facilities of the University Computing Center of
the City University of New York.

\newpage

\begin{flushleft}
{\bf Figure Captions}
\end{flushleft}

\noindent Figure  1.  Steady state order parameter density, $\rb_2 $,
{\it versus}
distance from critical point in the TTP.
(Main graph: log-log plot; inset: linear plot.)
\vspace{1em}

\noindent Figure  2.  Local slope analysis of the survival probability
data for various
$\phi_i$ values in the TTP.  $\delta$ is estimated from the $\tgi $ intercept.
\vspace{1em}

\noindent Figure  3.  Ultimate survival probability versus distance
from the critical point
in the TTP, for initial density $\phi_i = 0.4$.
\vspace{1em}

\noindent Figure  4.  Scaling plot of the survival probability in the DR.
+: $\D = 0.05$; $\times $ : $\D = 0.02$; $\Box $:
$\D = 0.01$; $\diamond $: $\D = 0.005 $;
$\circ $: $\D = 0.002 $.
Initial densities (top to bottom) $\phi_i = 1/2, 0.418, 0.38,$ and 1/3.
\vspace{1em}

\noindent Figure  5.  $\eta' $ {\it vs} $\delta '$ for the PCP,
DR, and TTP.  The slope of the best-fit
straight line is -0.995.

\newpage
{}~ \newpage

\begin{table}
\caption {Critical exponents of the threshold transfer process and the
dimer reaction.   Figures in parentheses denote uncertainties in the
last figure(s).}
\begin{center}
\begin{tabular}{|c|c|c|c|c|c|}  \hline
$\phi_i$ & $\delta '$ & $\eta '$ & $z'/2$ & $\beta '$ &
$ \eta '  + (1+ \frac{\beta}{\beta '})\delta '
- z'/2 $  \\ \hline
\multicolumn{6}{|c|}{Threshold Transfer Process} \\ \hline
$0.75$ & 0.136(1) & 0.347(4) & 0.632(7) & 0.250(5) &  0.00(1) \\
$0.69$ & 0.161(2) & 0.319(3) & 0.632(7) & 0.279(5) &  0.00(1) \\
$0.60$ & 0.192(2) & 0.288(3) & 0.630(7) & 0.356(5) &  0.00(1) \\
$0.50$ & 0.227(2) & 0.246(2) & 0.623(7) & 0.426(5) &  0.00(1) \\
$0.40$ & 0.270(3) & 0.204(2) & 0.622(7) & 0.497(5) &  0.00(1) \\
$0.31$ & 0.299(3) & 0.169(2) & 0.617(7) & 0.556(6) &  0.00(1) \\
$0.30$ & 0.303(3) & 0.170(2) & 0.621(7) & 0.567(6) &  0.00(1) \\
$0.25$ & 0.316(3) & 0.161(2) & 0.624(7) & 0.591(6) &  0.00(1) \\
$0.20$ & 0.342(3) & 0.133(1) & 0.622(7) & 0.640(6) &  0.00(1) \\
$0.10$ & 0.371(4) & 0.097(1) & 0.615(7) & 0.705(7) & 0.00(1) \\ \hline
\multicolumn{6}{|c|}{Dimer Reaction} \\ \hline
$0.333$ & 0.107(2) & 0.362(3) & 0.634(3) & 0.182(10) & 0.00(1) \\
$0.380$ & 0.133(2) & 0.327(3) & 0.629(5) & 0.241(6)  & -0.02(1) \\
$0.418$ & 0.158(2) & 0.302(4) & 0.626(3) & 0.275(2)  & -0.01(1) \\
$0.500$ & 0.205(5) & 0.250(5) & 0.620(3) & 0.357(10) & -0.01(1) \\ \hline
\end{tabular}
\end{center}

\end{table}

\end{document}